\documentclass[11pt,superscriptaddress]{revtex4}

\usepackage{picinpar}
\usepackage{floatflt}
\usepackage{array}
\usepackage{subfigure}
\usepackage{amsmath}
\usepackage[dvips]{graphicx}
\usepackage{amssymb}

\DeclareSymbolFont{operators}{OT1}{cmss}{m}{n}
\DeclareSymbolFont{letters}{OML}{cmss}{m}{it}
\DeclareSymbolFont{symbols}{OMS}{cmss}{m}{n}
\DeclareSymbolFont{largesymbols}{OMX}{cmss}{m}{n}

\begin{document}

\title{Vibrational Recognition of Adsorption Sites for Carbon Monoxide
\\ on Platinum and Platinum-Ruthenium Surfaces}

\author{Ismaila Dabo}
\affiliation{Department of Materials Science and Engineering,
Massachusetts Institute of Technology, Cambridge, MA, USA}
\author{Andrzej Wieckowski}
\affiliation{Department of Chemistry, 
University of Illinois at Urbana-Champaign, Urbana, IL, USA}
\author{Nicola Marzari}
\affiliation{Department of Materials Science and Engineering,
Massachusetts Institute of Technology, Cambridge, MA, USA}

\begin{abstract}
We have studied the vibrational properties of CO adsorbed on platinum and platinum-ruthenium
surfaces using density-functional perturbation theory within the Perdew-Burke-Ernzerhof
generalized-gradient approximation. The calculated C--O stretching frequencies are found to be in
excellent agreement with spectroscopic measurements. The frequency shifts that take place when the
surface is covered with ruthenium monolayers are also correctly predicted. This agreement for both
shifts and absolute vibrational frequencies is made more remarkable by the frequent failure of local
and semilocal exchange-correlation functionals in predicting the stability of the different
adsorption sites for CO on transition-metal surfaces. We have investigated the chemical origin of
the C--O frequency shifts introducing an orbital-resolved analysis of the force and frequency density
of states, and assessed the effect of donation and backdonation on the CO vibrational frequency
using a GGA + molecular U approach. These findings rationalize and establish the accuracy of
density-functional calculations in predicting absolute vibrational frequencies, notwithstanding the
failure in determining relative adsorption energies, in the strong chemisorption regime.
\end{abstract}

\maketitle

\section{Introduction}

Fuel cells are energy conversion systems
of potentially high environmental
benefit \cite{TesterDrake2005} that provide
electricity and heat by catalytic conversion of a fuel, such as hydrogen or
methanol. Despite their
advantages, several technological obstacles have hindered the deployment of
fuel-cell systems. For low-temperature
fuel cells that use platinum as electrode material, one major limitation is CO
poisoning,
whereby CO occupies active sites on the platinum catalyst and prevents fuel
oxidation \cite{LarminieDicks2003}. Typically, in
polymer electrolyte membrane fuel cells (PEMFCs), CO concentrations must be
brought below 10-50
ppm to maintain an acceptable catalytic performance. For comparison, CO
concentrations are generally
on the order of thousands of ppm in reformed hydrogen fuels 
\cite{VielstichLamm2003,BrandonThompsett2005}. CO poisoning is
even more problematic
for direct methanol fuel cells (DMFCs) since CO is always present in critical
amounts as an
intermediate in methanol oxidation \cite{Hoogers2003}.

Ruthenium islands on platinum catalysts have been shown to considerably
attenuate CO poisoning \cite{CrammFriedrich1997,HerreroFeliu1999,CrownJohnston2002},
although the microscopic details of this phenomenon are not completely
understood. Two main
mechanisms have been proposed to explain this improved tolerance to CO. Within
the bifunctional mechanism
model, adsorbed OH species generated by water dissociation at the
platinum/ruthenium
edge promote the oxidation of CO ({\it the promotion effect}) 
\cite{Hoogers2003,HerreroFeliu1999,CrownJohnston2002,TongKim2002,MaillardLu2005,
LiuLogadottir2003}. According to an
alternative view,
ruthenium modifies the electronic structure of neighboring platinum atoms,
reducing their affinity for
CO ({\it the ligand/intrinsic effect}) \cite{Hoogers2003,LiuLogadottir2003}.
To investigate further these mechanisms of
central interest to fuel-cell technology, it is necessary to elucidate the
nature of the chemical interaction between CO and bimetallic
surfaces.

In most cases, density-functional theory provides a reliable description of
molecular adsorption and
dissociation on transition metals \cite{HammerMorikawa1996,
HammerNorskov1995,NorskovBligaard2002,GreeleyMavrikakis2004,ReuterFrenkel2004,
HammerHansen1999}. However, CO adsorption on
transition-metal surfaces is
unexpectedly problematic. Indeed, at low CO coverage, local and
generalized-gradient density-functional
calculations predict CO adsorption on Pt(111) to take place at the fcc site,
contradicting low-temperature
experiments, which unambiguously indicate atop adsorption. This well-known
qualitative
discrepancy (the ``CO/Pt(111) puzzle'') \cite{FeibelmanHammer2001}
 precludes an accurate description of
important phenomena,
such as the surface diffusion of CO adsorbates and the thermal population of CO
adsorption sites.
Similar qualitative errors have been reported for CO adsorbed on rhodium and
copper surfaces \cite{FeibelmanHammer2001,GajdosHafner2005,KohlerKresse2004}, and
a wide body of literature exists on the subject 
\cite{KresseGil2003,GrinbergYourdshahyan2002,MasonGrinberg2004,Gronbeck2004,
Doll2004,NeefDoll2006,WuVargas2001,OritaItoh2004,OlsenPhilipsen2003,
BirgerssonAlmbladh2003,GreeleyGokhale2003,MavrikakisRempel2002}.

In this work, we highlight and rationalize the accuracy of density-functional
calculations in predicting
the stretching frequencies of CO adsorbed on platinum and platinum-ruthenium
surfaces,
notwithstanding the failure in predicting the most stable adsorption site. We
first present density-functional
theory and density-functional perturbation theory results for the energetic,
structural and
vibrational properties of adsorbed CO. Second, we introduce a novel
orbital-resolved force analysis to
clarify the electronic origins of the C--O frequency shifts as a function of the
adsorption site. Last, we
rationalize the accuracy of the stretching-frequency predictions by analyzing
the influence of donation
and backdonation using a GGA + molecular U model recently introduced by Kresse,
Gil, and Sautet \cite{KresseGil2003}.

\section{Theoretical Basis}

The (111) transition-metal surface is modeled using a periodically repeated slab
composed of four
layers, each layer containing four atoms per supercell. A $\sqrt{3}$ $\times$ 2 adsorption
structure corresponding to a
coverage of 1/4 of the monolayer (ML) is adopted for the CO overlayer. Atomic cores are
represented by
ultrasoft pseudopotentials \cite{Vanderbilt1990}.
The
exchange-correlation energy is
calculated within the Perdew-Burke-Ernzerhof generalized-gradient approximation
(PBE-GGA) \cite{PerdewBurke1996}. The
size of the vacuum region separating the periodic slabs is $\sim$13 \AA. We use a
shifted 4 $\times$ 4 $\times$ 1 mesh with
cold-smearing occupations \cite{Marzari1996} (smearing temperature of 0.4 eV) to sample the
Brillouin zone. Energy
cutoffs of 24 and 192 Ry are applied to the plane-wave expansions of the
wavefunctions and charge
density, respectively. As discussed in Ref. \cite{FeibelmanHammer2001}, 
the system is not
spin-polarized. Using the above
slab thickness and calculation parameters, we verify that the adsorption
energies are converged within
less than 10 meV and the atomic forces within a few meV/\AA.

The bond length and stretching frequency of CO in the gas phase are calculated
to be 1.140 \AA\ and
2140 cm$^{-1}$ (experimental values are 1.128 \AA\ and 2170 cm$^{-1}$). The PBE-GGA lattice
parameter and bulk
modulus of platinum are 3.993 \AA\ and 2.36 Mbar, in good agreement with
experimental values of 3.923 \AA\ and 2.30 Mbar \cite{CRC2007}. 
All our calculations use fully relaxed configurations.

\section{Results}

\begin{table*}
\begin{center}
\small
\begin{tabular}{lllll}
\hline \hline
site  &  atop & bridge & hcp & fcc          \\
\hline
$E_{\rm ads}$ (eV) & 1.61 & 1.71 & 1.72 & 1.74 \\
& (1.30)\footnotemark[2] & & & \\
$d(\textrm{C--O})$ (\AA) & 1.153  & 1.177 & 1.188 & 1.189 \\
& (1.15$\pm$0.05)\footnotemark[3] & (1.15$\pm$0.05)\footnotemark[3] & & \\
$d(\textrm{M--C})$ (\AA) & 1.864 & 2.029 & 2.116 & 2.121 \\
& (1.85$\pm$0.1)\footnotemark[3]  & (2.08$\pm$0.07)\footnotemark[3] & & \\
$h(\rm C)$ (\AA)\footnotemark[1] & 2.017 & 1.543 & 1.380 & 1.373 \\
$\theta(\rm CO)$ (deg)\footnotemark[1] & 1.4 & 1.4 & 0.5 & 0.4 \\
$\nu(\textrm{C--O})$ (cm$^{-1}$)  & 2050 & 1845 & 1752 & 1743 \\
& (2070)\footnotemark[4] & (1880)\footnotemark[4] & (1760)\footnotemark[4] & (1760)\footnotemark[4] \\
$\nu(\textrm{M--C})$ (cm$^{-1}$) & 584 & 413 & 358 & 344 \\
& (470)\footnotemark[5]   & (380)\footnotemark[5] & & \\
bending modes & 392 & 393 & 329 & 328 \\
(cm$^{-1}$)   & 386 & 346 & 315 & 300 \\
other modes & 0 to 230  & 0 to 231 & 0 to 196  & 0 to 186 \\
(cm$^{-1}$) & & & & \\
\hline \hline
\end{tabular}
\end{center}
\small
\footnotemark[1]{$h(\rm C)$ denotes the distance from C to the first surface layer,
and $\theta(\rm CO)$ denotes the tilt angle of CO.}
\footnotemark[2]{Ref. \cite{LuLee2002}.}
\footnotemark[3]{Ref. \cite{OgletreeVanhove1986}.}
\footnotemark[4]{Ref. \cite{LuWhite2004}.}
\footnotemark[5]{Ref. \cite{SteiningerLehwald1982}.}
\caption[Adsorption energies, structural properties, and vibrational frequencies
for CO on clean platinum surfaces]
{Adsorption energies, structural properties, and vibrational frequencies calculated
using density-functional theory and density-functional perturbation theory for CO
adsorbed on clean Pt(111) surfaces.
\label{COPlatinumTable}
}
\end{table*}

\begin{table*}
\begin{center}
\small
\begin{tabular}{lllllll}
\hline \hline
site  &  atop & hcp & fcc & atop & hcp & fcc \\
\hline
slab & 1 ML Ru/ & 1 ML Ru/ & 1 ML Ru/ & 2 ML Ru/ & 2 ML Ru/ & 2 ML Ru/ \\
& 3 ML Pt & 3 ML Pt & 3 ML Pt & 2 ML Pt & 2 ML Pt & 2 ML Pt \\
$E_{\rm ads}$ (eV) & 2.24 & 2.27 & 2.15 & 1.96 & 2.05 & 1.88 \\
$d(\textrm{C--O})$ (\AA) & 1.161 & 1.196 & 1.192 & 1.161 & 1.201 & 1.190  \\
$d(\textrm{M--C})$ (\AA) & 1.892 & 2.132 & 2.104 & 1.922 & 2.122 & 2.111 \\
$h(\rm C)$ (\AA)\footnotemark[1] & 2.014 & 1.375 & 1.412 & 1.970 & 1.318 & 1.437 \\
$\theta(\rm CO)$ (deg)\footnotemark[1] & 3.3 & 0.9 & 2.6 & 4.1 & 0.6 & 1.44 \\
$\nu(\textrm{C--O})$ (cm$^{-1}$)  & 1979 & 1702 & 1724 & 1969 & 1666 & 1739 \\
& (1970)\footnotemark[2] & & & (1970)\footnotemark[2] & & \\
$\nu(\textrm{M--C})$ (cm$^{-1}$) & 510 & 356 & 358 & 482 & 355 & 351 \\
bending modes & 412 & 258 & 221 & 396 & 315 & 251 \\
(cm$^{-1}$)   & 409 & 247 & 215 & 389 & 301 & 231 \\
other modes & 0 to 208  & 0 to 205 & 0 to 201  & 0 to 265 & 0 to 254 & 0 to 226 \\
(cm$^{-1}$) & & & & \\
\hline \hline
\end{tabular}
\end{center}
\small
\footnotemark[1]{$h(\rm C)$ denotes the distance from C to the first surface layer,
and $\theta(\rm CO)$ denotes the tilt angle of CO.}
\footnotemark[2]{Ref. \cite{LuWhite2004}.}
\caption[Adsorption energies, structural properties, and vibrational frequencies
for CO on ruthenium-covered platinum surfaces]
{Adsorption energies, structural properties, and vibrational frequencies calculated
using density-functional theory and density-functional perturbation theory for CO
adsorbed on ruthenium-covered Pt(111) surfaces.
\label{CORutheniumTable}
}
\end{table*}

We report the results of our density-functional calculations in Tables 
\ref{COPlatinumTable} and \ref{CORutheniumTable}.
For platinum surfaces,
the calculated atop binding energy $E_{\rm ads}(\rm atop)$ = 1.61 eV is consistent with that
reported in Ref. \cite{SteckelEichler2003}
(1.55 eV in the same adsorption structure). As a matter of comparison, the
experimental heat of
adsorption at 1/4 ML CO is 1.30 eV. The relative adsorption energy
$E_{\rm ads}({\rm atop})-E_{\rm ads}({\rm fcc})$ is calculated to
be $−0.13$ eV, in accordance with the gradient-corrected relative adsorption
energies (ranging from $-0.10$
to $-0.25$ eV) reported in Ref. \cite{FeibelmanHammer2001}. As expected, our density-functional
calculations favor CO
adsorption at the threefold fcc and hcp adsorption sites for platinum and
platinum-ruthenium surfaces,
confirming the aforementioned disagreement with experiments. (Note that bridge
adsorption of CO on
platinum-ruthenium surfaces is predicted to be energetically unstable.) Despite
this noteworthy failure,
the bond length $d(\textrm{C--O})$ is calculated to be 1.153 \AA\ at the atop site and 1.177 \AA\
at the bridge site on
platinum, in good agreement with experimental bond lengths (1.15$\pm$0.05 \AA\ at both
the atop and fcc
sites). Similarly, the distance $d(\textrm{Pt--C})$ from the carbon to its nearest platinum
neighbor, calculated to be
1.864 \AA\ at the atop site and 2.029 \AA\ at the fcc site, is always within
experimental error (experimental
bond lengths are 1.85$\pm$0.1 \AA\ and 2.08$\pm$0.07 \AA\ at the atop and fcc sites,
respectively). Note that both
bond lengths increase with coordination.

The full phonon spectra for CO adsorbed at the atop, bridge, hcp, and fcc sites
on platinum and
platinum-ruthenium surfaces are calculated using density-functional perturbation
theory (DFPT) \cite{BaronideGironcoli2001}.
Within this approach, the full dynamical matrix of the system is computed
exactly by solving the self-consistent
linear-response problem describing the electron response to atomic perturbations
of arbitrary
wavelength. The DFPT spectra reported in Tables \ref{COPlatinumTable} and 
\ref{CORutheniumTable} exhibit some common and
expected features.
The highest vibrational frequency in the range [1700 cm$^{-1}$, 2100 cm$^{-1}$]
corresponds to the localized
C--O stretching mode. The second highest frequency $\nu(\textrm{M--C})$ 
in the range [300 cm$^{-1}$, 600 cm$^{-1}$] is
related to the stretching of the metal-carbon bond. This mode is followed by two
CO bending modes
with frequencies lying 20-200 cm$^{-1}$ below $\nu(\textrm{M--C})$. 
All the other modes involving displacements of the
heavy metal atoms are found in the frequency range [0 cm$^{-1}$, 300 cm$^{-1}$].

We now focus on the dependence of the C--O stretching frequency as a function of
the adsorption site.
Upon atop adsorption on platinum, the predicted $\nu(\textrm{C--O})$ 
is reduced from 2140 cm$^{-1}$ to 2050 cm$^{-1}$,
corresponding to a red shift $\Delta\nu(\textrm{C--O})$ of 
$-90$ cm$^{-1}$. For comparison, the
experimental stretching
frequency, as obtained by means of sum-frequency generation (SFG)
spectroscopy \cite{LuWhite2004}, decreases from
2170 cm$^{-1}$ to 2070 cm$^{-1}$, 
corresponding to $\Delta\nu(\textrm{C--O})$ = $-100$ cm$^{-1}$. The frequency
shifts are even more
marked at high-coordination sites: $\nu(\textrm{C--O})$ is predicted to be 1845 cm$^{-1}$, 
1752 cm$^{-1}$, and 1743 cm$^{-1}$ at
the bridge, hcp, and fcc sites, corresponding to red shifts of up to $-397$ cm$^{-1}$.
These DFPT stretching
frequencies show remarkable agreement with their SFG counterparts: $\nu(\textrm{C--O})$ = 1830
cm$^{-1}$ at the twofold
bridge site, $\nu(\textrm{C--O})$ = 1760 cm$^{−1}$ at the threefold hcp and fcc sites,
corresponding to a maximum
red shift of $-410$ cm$^{-1}$. Accurate DFPT frequencies are also obtained for CO
adsorbed on platinum-ruthenium
bimetallic surfaces. Indeed, the calculated stretching frequencies 1979 cm$^{-1}$ (1
Ru ML) and
1969 cm$^{-1}$ (2 Ru ML) at the atop site compare very closely to the SFG result of
1970 cm$^{-1}$.

In conclusion, all calculated C−O stretching frequencies deviate by less than 2\%
from the measured
ones, irrespective of the adsorption site and nature of the metal surface. The
correct prediction of the
frequency red shifts allows the direct recognition of CO adsorption sites and
confirms that CO
preferentially occupies atop sites on platinum-ruthenium bimetallic surfaces.
This very close agreement
with experiment is made more remarkable by the lack of accuracy of the PBE-GGA
adsorption
energies. In the remainder of this work, we show how this accuracy can be
rationalized in terms of the
hybridization of the CO molecular orbitals with the metal bands.

\section{Discussion}

\subsection{Electronic Origins of the Frequency Shifts}

\label{ElectronicOriginSection}

\begin{figure}
\begin{center}
\includegraphics[height=20cm]{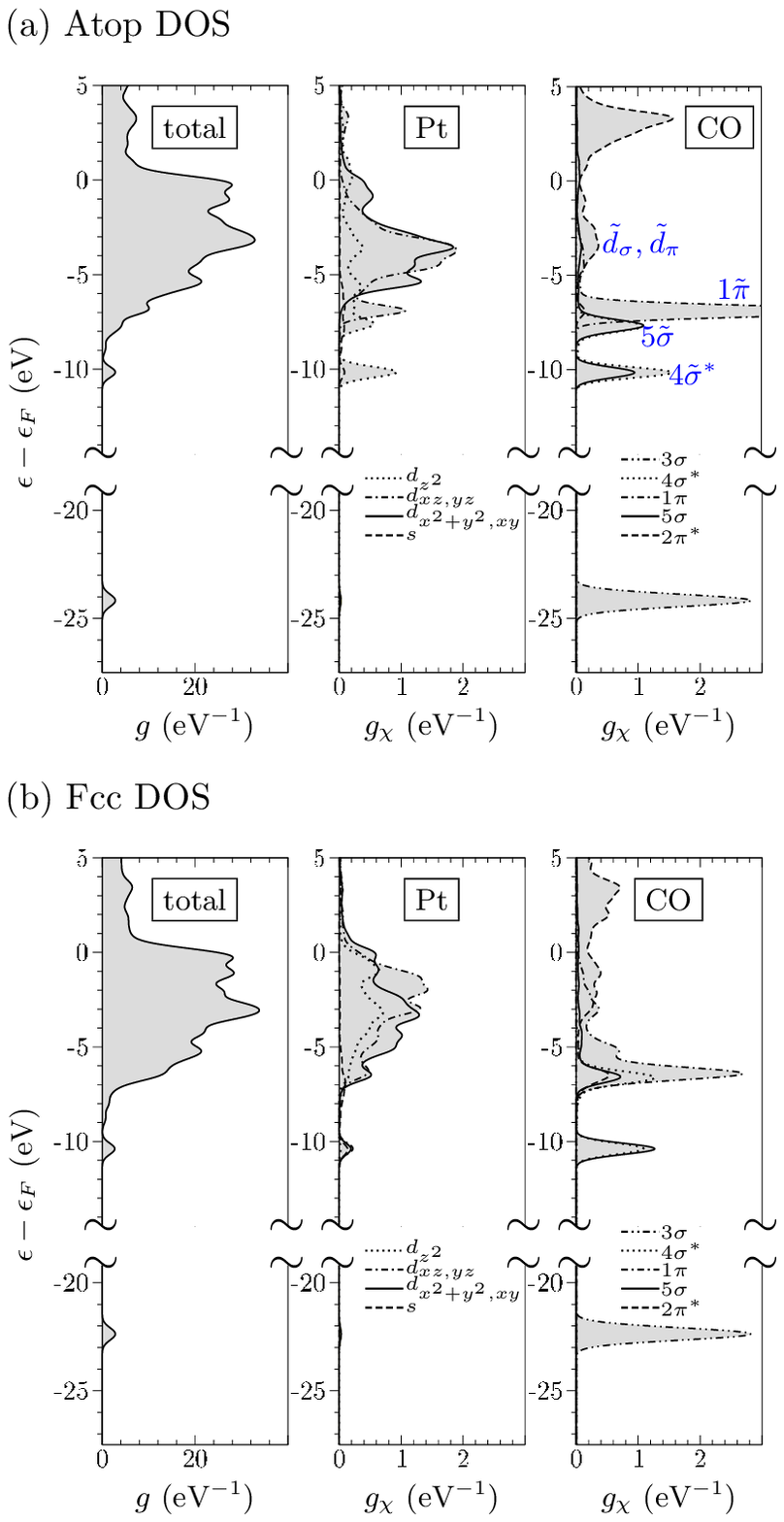}
\caption[Density of states for atop and fcc adsorption of CO on platinum]{
Total density of states, density of states projected on the Pt atomic
orbitals, and density of states projected on the CO molecular orbitals
for atop and fcc adsorption of CO on Pt(111).
\label{COPlatinumDOS}}
\end{center}
\end{figure}

\begin{figure}
\begin{center}
\includegraphics[height=20cm]{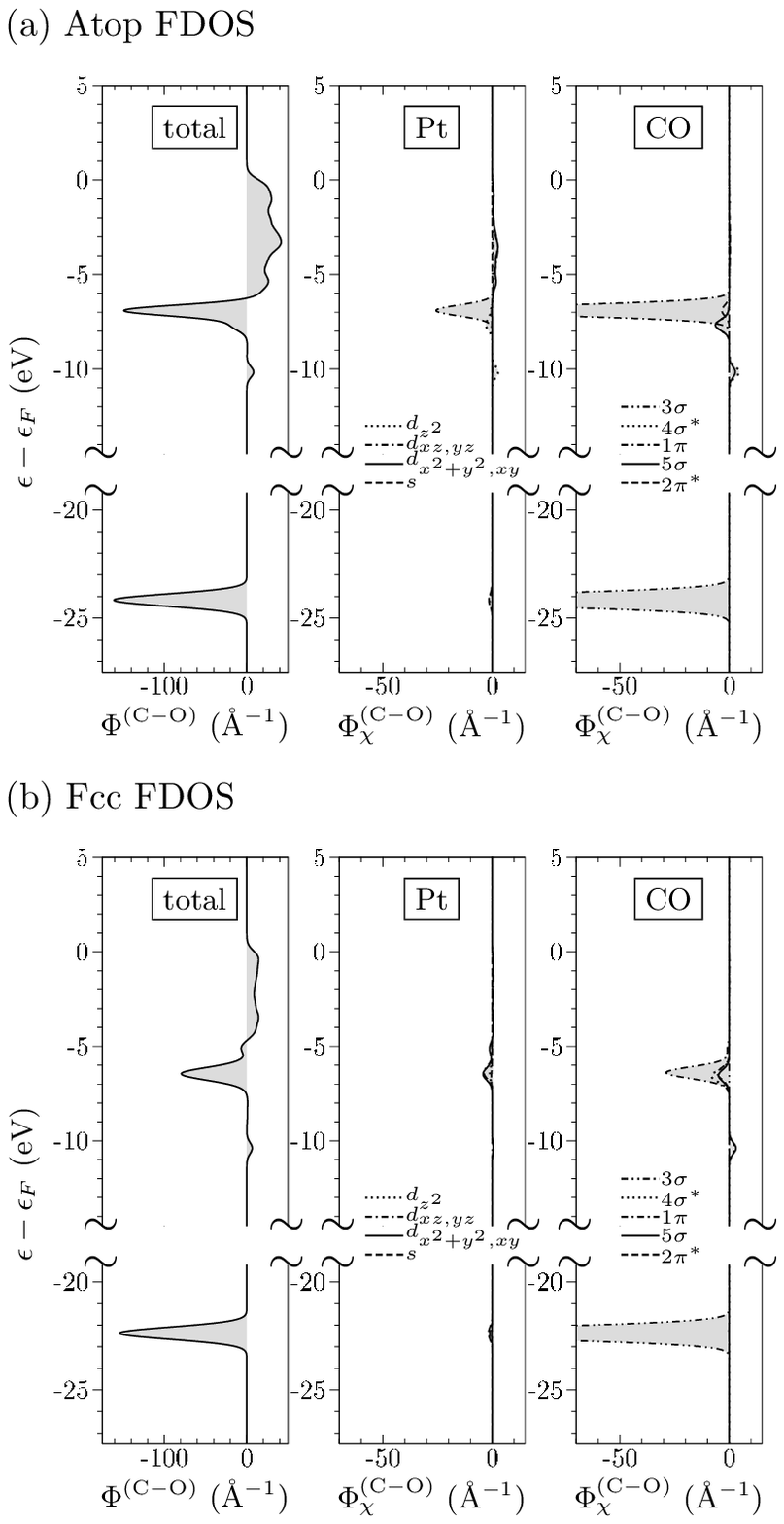}
\caption[Force density of states for atop and fcc adsorption of CO on platinum]{
Total force density of states, force density of states projected on the Pt atomic
orbitals, and force density of states projected on the CO molecular orbitals
for atop and fcc adsorption of CO on Pt(111).
\label{COPlatinumFDOS}}
\end{center}
\end{figure}

\begin{figure}
\begin{center}
\includegraphics[width=13cm]{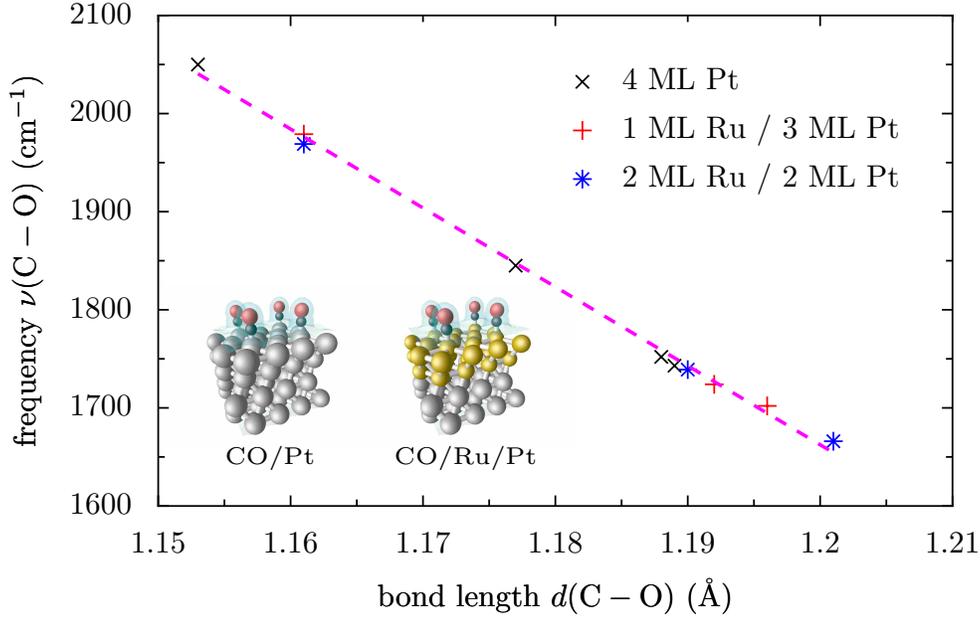}
\caption[Correlation between bond length and intramolecular frequency
for CO on clean and ruthenium-covered platinum surfaces]{
Correlation between bond length and intramolecular frequency
for CO on clean and ruthenium-covered Pt(111) surfaces.
\label{CorrelationNuD}}
\end{center}
\end{figure}

\begin{table}
\begin{center}
\small
\begin{tabular}{llllll}
\hline \hline
 & gas phase & atop & bridge & hcp & fcc \\
\hline
$f_{2 \pi^*_x}/2+f_{2 \pi^*_y}/2$& 0.00 & 0.25 & 0.37 & 0.41 & 0.41 \\
$f_{5 \sigma}$ &  1.00 & 0.92 & 0.92 & 0.92 & 0.92 \\
$f_{1 \pi_x}/2+f_{1 \pi_y}/2$ & 1.00 & 1.00 & 0.99 & 0.99 & 0.99 \\
$f_{4 \sigma^*}$ & 1.00 & 1.00 & 1.00 & 1.00 & 1.00 \\
$f_{3 \sigma}$ & 1.00 & 1.00 & 1.00 & 1.00 & 1.00 \\
bond order & 3.00 & 2.35  & 2.16 & 2.09 & 2.09 \\
\hline \hline
\end{tabular}
\end{center}
\caption[Molecular occupations and bond order for CO in the gas phase 
and adsorbed on platinum]{
Molecular occupations and bond order for CO in the gas phase and for CO adsorbed on platinum.
\label{OrbitalOccupationsTable}
}
\end{table}

The hybridization of the metal d bands with the $2\pi^*$ lowest unoccupied molecular
orbitals (LUMOs)
and the $5\sigma$ highest occupied molecular orbital (HOMO) plays a predominant role in
the adsorption
energy of CO on transition metals. According to the Blyholder model \cite{Blyholder1964}, these
electronic interactions
result in electron donation (i.e., partial depletion of the $5\sigma$ orbital) and
electron backdonation (i.e.,
partial filling of the $2\pi^*$ orbital). It has been shown that the trends of the
adsorption energies of CO on
transition-metal surfaces can be correlated to the amounts of donation and
backdonation (Hammer-Morikawa-N{\o}rskov model) \cite{HammerMorikawa1996}. 
Among the suggested solutions to the ``CO/Pt(111)
puzzle''---e.g.,
nonequivalent GGA description of different bond orders \cite{GrinbergYourdshahyan2002}, 
incorrect singlet-triplet CO excitation
energies \cite{MasonGrinberg2004}, effect of metal semicore polarization
\cite{Gronbeck2004}---Kresse, Gil, and Sautet have
proposed that the
inaccuracy of density-functional calculations in determining the most stable
adsorption site is due to an
overestimation of the interaction between the $2\pi^*$ orbitals and the metal bands,
resulting from an
underestimation of the HOMO-LUMO gap \cite{KresseGil2003}. 
As discussed in the next section, this
interpretation
recovers the essential features of CO adsorption on transition metals: it
identifies the tendency of local
and generalized-gradient DFT to delocalize and overhybridize electronic states.
Nevertheless, as shown
below, the site-dependence of the C--O bond length and vibrational frequency is
not affected by the
LUMO and HOMO hybridizations. In other words, the hybridizations that subtly
determine the relative
CO adsorption energies do not influence the structural and vibrational
predictions.

To establish this fact, we first introduce a spectral force analysis. The main
objective of this analysis
is to separate and assess the contribution from each CO molecular orbital to the
force F acting on a
given atom. The central quantity we introduce is $\Phi_{I,\chi}(\epsilon)$, 
the force density of states (FDOS) of the orbital
$\chi$, which is defined as the $\chi$-resolved density of states weighted by the
wavefunction contribution to the
force ${\bf F}_I$ acting on atom $I$. To be more explicit, the FDOS of a given CO molecular
orbital $\chi$ can be expressed as:
\begin{equation}
\Phi_{I,\chi}(\epsilon) = \sum_i {\bf F}_{I,i} |\langle \chi | \psi_i \rangle |^2 
\delta(\epsilon-\epsilon_i),
\end{equation}
where $\psi_i$ denotes the electronic wavefunction, $\epsilon_i$ is 
the electronic energy, and \linebreak ${\bf F}_{I,i}= - f_i \int |\psi_i|^2 \partial v/\partial 
{\bf R}_I$ is the wavefunction contribution to the force ${\bf F}_I$. 
(The calculation of the overlap $\langle \chi | \psi_i \rangle$ in the ultrasoft
formalism is detailed in Appendix \ref{UltrasoftAppendix}.) It should be noted
that, by summing the
integrated FDOS $\Phi_{I,\chi}(\epsilon)$ over a complete set of orbitals satisfying
orthonormality, one obtains the total
electronic force acting on atom $I$. As a consequence, the FDOS can be
quantitatively connected to
relevant observables. Additionally, by projecting the force density of states
along the normalized atomic
displacements $\Delta_I^{\textrm{C--O}}$ 
corresponding to the C--O stretching mode, we obtain the force density of
states along the stretching mode $\Phi_\chi^{\textrm{C--O}}(\epsilon)=\sum_I 
\Delta_I^{\textrm{C--O}} \cdot \Phi_{I,\chi}(\epsilon)$, 
to be heuristically identified as the orbital
contribution to the intramolecular force.

The orbital-projected density of states (DOS) $g_\chi(\epsilon) = 
\sum_i |\langle \chi | \psi_i \rangle |^2 \delta(\epsilon-\epsilon_i)$
is commonly used to
provide an insightful picture of the electronic hybridizations that take place
when CO is adsorbed on
platinum. Similarly, $\Phi_\chi^{\textrm{C--O}}(\epsilon)$
describes the influence of electronic hybridizations on the force along the
C--O stretching mode. The projected densities of states $g_\chi(\epsilon)$
and projected force densities of states
$\Phi_\chi^{\textrm{C--O}}(\epsilon)$ for different adsorption sites are plotted in 
Figures \ref{COPlatinumDOS} and \ref{COPlatinumFDOS}. A detailed
analysis of the orbital-resolved
densities of states is given in Ref. \cite{KresseGil2003}. 
For the purpose of our study, we
emphasize the
following features. When CO adsorbs on Pt(111), the 4$\sigma^*$ and $5\sigma$ orbitals
hybridize with the metal d$_{z^2}$
band, generating $4\tilde \sigma^*$ and $5 \tilde \sigma$ states 
with mainly adsorbate character
(adopting the terminology of
Ref. \cite{FohlischNyberg2000}, the tilde symbol denotes hybrid states). These
$4\tilde \sigma^*$ and $5 \tilde \sigma$
states are found in the energy
ranges [$-12$ eV, $-9$ eV] and [$-9$ eV, $-5$ eV] relative to the Fermi level. Above $-5$
eV, the $5\sigma$ orbital
and the d$_{z^2}$ band generate a $\tilde {\rm d}_\sigma$ band with predominant metal character. This interaction
results in a
partial depletion of the $5\sigma$ HOMO (electron donation). In addition, the
interaction between the $1\pi$ and
$2\pi^*$ orbitals and the d$_{xz}$ and d$_{yz}$ bands produces $1\tilde \pi$ 
states in the range [$-9$
eV, $-5$ eV] and a broad $\tilde {\rm d}_\pi$
band above $-5$ eV, causing partial occupation of the $2\pi^*$  LUMOs (electron
backdonation) \cite{Blyholder1964}. The
changes in molecular orbital occupations due to CO adsorption are reported in
Table \ref{OrbitalOccupationsTable}.

In order to understand how the generation of these hybrid states affects the
intramolecular force, we
turn to the FDOS (Figure \ref{COPlatinumFDOS}). 
The graphs are plotted according to the convention
that bonding states
(i.e., opposed to the stretching of the C--O bond) correspond to negative values
of $\Phi_\chi^{\textrm{C--O}}$. First, we
note that the bonding contribution from the $3\sigma$ state does not vary with the
adsorption site, confirming
that the $3 \sigma$ state retains a strong molecular character. Additionally, we
observe that the $5 \tilde \sigma$ and $1 \tilde \pi$
states are bonding while the $4 \tilde \sigma^*$ 
is antibonding, as expected intuitively. In
the energy region above $-5$
eV, another contribution appears. This contribution corresponds to high-energy
wavefunctions located
inside the platinum slab, as evidenced by the absence of any molecular-orbital
force contribution above
$-5$ eV. Nevertheless, due to their metal character, the contribution of these
high-energy wavefunctions
is mostly canceled by the positively charged platinum cores. Consequently, the
local contribution from
the hybrid states of strong molecular character prevails.

Besides these observations, the main feature of the FDOS graphs is the
predominant bonding
contribution between $-9$ eV and $-5$ eV. At the atop site, the curve displays a
sharp negative peak which
corresponds mainly to the $1\pi$ orbital-resolved contribution 
$\Phi_{1\pi}^{\textrm{C--O}}(\epsilon)$. At the fcc site, both the
magnitude and the relative share of the peak are reduced, clearly indicating
that the $1 \tilde \pi$ states have more
influence on the change in intramolecular bonding than any of the other hybrid
wavefunctions. The $1 \tilde \pi$
states maintain a predominant $1\pi$ character at the atop site, whereas at
high-coordination sites this
molecular character is significantly reduced due to a stronger hybridization
with the substrate.
Therefore, the $1\pi$ bonding contribution to the intramolecular force decreases
with site coordination. As
intramolecular bonding decreases, the C--O bond length increases. The
predominance of the $1 \pi$ bonding
contribution is confirmed by the density-distribution analysis initially
introduced by Zupan, Burke,
Ernzerhof, and Perdew \cite{ZupanBurke1997}, as discussed in Appendix 
\ref{ZupanAnalysisAppendix}.

For CO adsorbed on transition metal surfaces, the intramolecular
bond length and the intramolecular stretching frequency
are strongly correlated. An
extensive study of Gajdo\v{s}, Eichler, and Hafner \cite{GajdosEichler2004} 
showed a linear correlation
between $d(\textrm{C--O})$ and
$\nu(\textrm{C--O})$ for CO adsorbed on close-packed transition metals:
$\nu(\textrm{C--O})$ shifts down in
frequency as $d(\textrm{C--O})$ increases. 
As illustrated in Figure \ref{CorrelationNuD}, 
a similar trend is observed for CO adsorbed on ruthenium-covered platinum
surfaces.
Therefore, the increase in C--O bond length at
high-coordination sites, which reflects
a decrease in $1\pi$ bonding contribution, is accompanied by a reduction of the C--O
stretching frequency.

While the preceding is consistent with the interpretation given in Refs
\cite{Blyholder1964,FohlischNyberg2000,GajdosEichler2004}, it is
important to make one central observation: although the LUMO $2\pi^*$ filling is a
reasonable measure of
the amount of hybridization between the $1\pi$, $2\pi^*$ orbitals and the metal d$_{xz}$,
d$_{yz}$ bands, filling the $2\pi^*$
orbitals does not directly weaken the bond, as evidenced by the very low values
of $\Phi^{\textrm{(C--O)}}_{2\pi^*}$ in the
energy range [$-9$ eV, $-5$ eV]. This interpretation helps explain the fact that
the CO adsorption
energies do not show a well-defined relationship with the C--O stretching
frequency \cite{ShubinaKoper2002,HanCeder2006}.

The main conclusion of this section is as follows. At variance with the CO
adsorption energies,
electron backdonation and electron donation have little direct bearing on the
intramolecular forces.
Their immediate effect on the molecular bond length and stretching frequency
cannot account for the
observed shifts. Instead, the changes in bond length and stretching frequency
are primarily related to the
hybridization of the $1\pi$ molecular orbitals. This provides important indications
as to why the structural
and vibrational properties of CO adsorbed on platinum and platinum-ruthenium
surfaces are accurately
predicted. The GGA + molecular U study presented in the next part provides
additional quantitative
evidence in support to this conclusion.

\subsection{Influence of Donation and Backdonation on the Accuracy of the Frequency
Predictions}

As mentioned above, the failure of density-functional calculations in predicting
CO adsorption
energies is traceable to an overhybridization of the CO molecular orbitals with
the metal bands \cite{KresseGil2003,GilClotet2003}. To
assess the influence of this overhybridization on the accuracy of the calculated
adsorption energies,
bond lengths, and vibrational properties, we have performed a sensitivity
analysis. This analysis consists
of controlling and varying the HOMO and LUMO hybridizations, while monitoring
the variations of the
mentioned observables. To this end, we have used the GGA + molecular U approach
introduced by
Kresse, Gil, and Sautet \cite{KresseGil2003}. 
This approach (inspired by the LDA + U method \cite{AnisimovAryasetiawan1997})
consists of adding an
orbital-dependent term to the GGA energy functional, thus imposing a penalty on
orbital hybridization.

\begin{figure}
\begin{center}
\includegraphics[height=20cm]{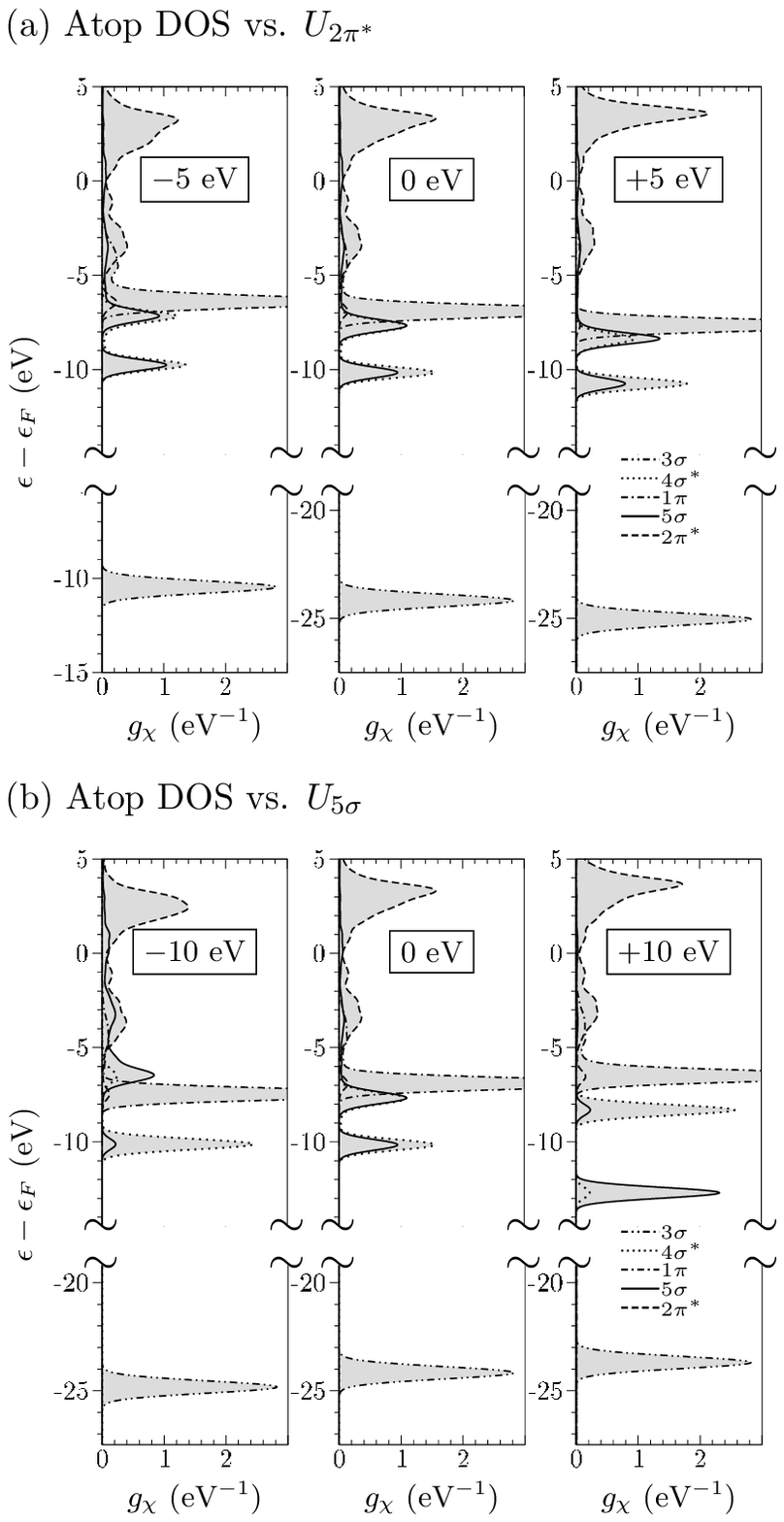}
\caption[Density of states 
as a function
of the hybridization parameters
for atop adsorption of CO on platinum]{
Density of states projected on the CO molecular orbitals as a function
of $U_{2\pi^*}$ and $U_{5\sigma}$ controlling the hybridization of
the LUMO and HOMO orbitals for atop adsorption of CO on Pt(111).
\label{COPlatinumUDOS}}
\end{center}
\end{figure}

We employ the following GGA + molecular U energy functional:
\begin{equation}
E_{GGA+U}= E_{GGA} + \frac{U_{2 \pi^*}}{2}
\sum_\sigma \textrm{Tr} \{ {\bf f}_{2 \pi^*, \sigma} ({\bf I}- {\bf f}_{2 \pi^*, \sigma}) \} \nonumber + \frac{U_{5 \sigma}}{2} \sum_\sigma f_{5 \sigma, \sigma} (1 - f_{5 \sigma, \sigma}),
\label{GGAmolecularUFunctional}
\end{equation}
where $f_{5 \sigma, \sigma}$ is the occupation of the 
$5\sigma$ orbital of spin $\sigma$ and ${\bf f}_{2 \pi^*, \sigma}$ is the
occupation matrix associated
with the $2\pi_x^*$ and $2\pi_y^*$ orbitals of spin $\sigma$. 
The parameters $U_{2\pi^*}$ and $U_{5\sigma}$
penalize noninteger
occupations of the $2\pi^*$ and $5\sigma$ orbitals: 
$U_{2\pi^*}$ reduces $2\pi^*$ backdonation while
$U_{5\sigma}$ reduces $5\sigma$
donation, as illustrated in Figure \ref{COPlatinumUDOS}. The parameters 
$U_{2\pi^*}$ and $U_{5\sigma}$ can also be
interpreted as shifting the
effective single-electron energies. Heuristically, $U_\chi$ modifies the
single-electron energy $\epsilon_\chi$ by an
amount $U_\chi(1/2-f_\chi)$. Thus, $U_{2\pi^*}$ increases the 
$2\pi^*$  energies, whereas $U_{5\sigma}$
decreases the $5\sigma$
energy, causing the HOMO-LUMO gap to increase. 

The present functional differs slightly from that introduced by Kresse, Gil, and
Sautet. The GGA + molecular U energy in Eq.
\ref{GGAmolecularUFunctional}, whose expression is based on the matrix formulation
introduced by Cococcioni and de Gironcoli \cite{CococcionideGironcoli2005}, 
is invariant with respect to the choice of the $x$-
and $y$-axes. In other
words, an arbitrary rotation of the molecular orbitals does not affect the GGA +
molecular U energy.
Additionally, the functional allows the freedom to vary both the amount of
electron backdonation and
that of electron donation. The necessity of simultaneously varying backdonation
and donation will be
discussed later.

Although a molecular U term is admittedly a simplified energy correction, it
reproduces the essential
features of the energetics of CO adsorption \cite{KresseGil2003}. 
The cluster calculations of Gil
{\it et al.}, based on the B3LYP
hybrid functional \cite{Becke1993}, 
confirm that the inaccuracy of the GGA energies can be
ascribed to an
overhybridization of the $2\pi^*$ molecular orbitals \cite{GilClotet2003}. 
This conclusion is supported
by the recent periodic-slab
B3LYP calculations of Neef and Doll \cite{Doll2004,NeefDoll2006}. 
Moreover, experimental studies
indicate that the
adsorption energy of CO on platinum shows a linear dependence with respect to
the energy of the center
of the metal d bands,35 in agreement with the theoretical model developed by
Hammer, Morikawa and
N{\o}rskov \cite{HammerMorikawa1996}. 
However, the coefficient of proportionality is overestimated within
density-functional
calculations, indicating that the interaction between the $2\pi^*$ orbitals and the
metal d bands is excessive.

We thus proceeded to calculate the energetic, structural, and vibrational
properties for CO adsorbed on
platinum. Stretching frequencies are now obtained by diagonalizing the
two-by-two dynamical matrix
associated off-equilibrium displacements of the carbon and oxygen atoms in the
direction normal to the
surface. Due to the large atomic mass of platinum, the resulting stretching
frequencies deviate by less
than 1 cm$^{-1}$ from the full DFPT phonon frequencies.

\begin{figure}
\begin{center}
\includegraphics[width=16cm]{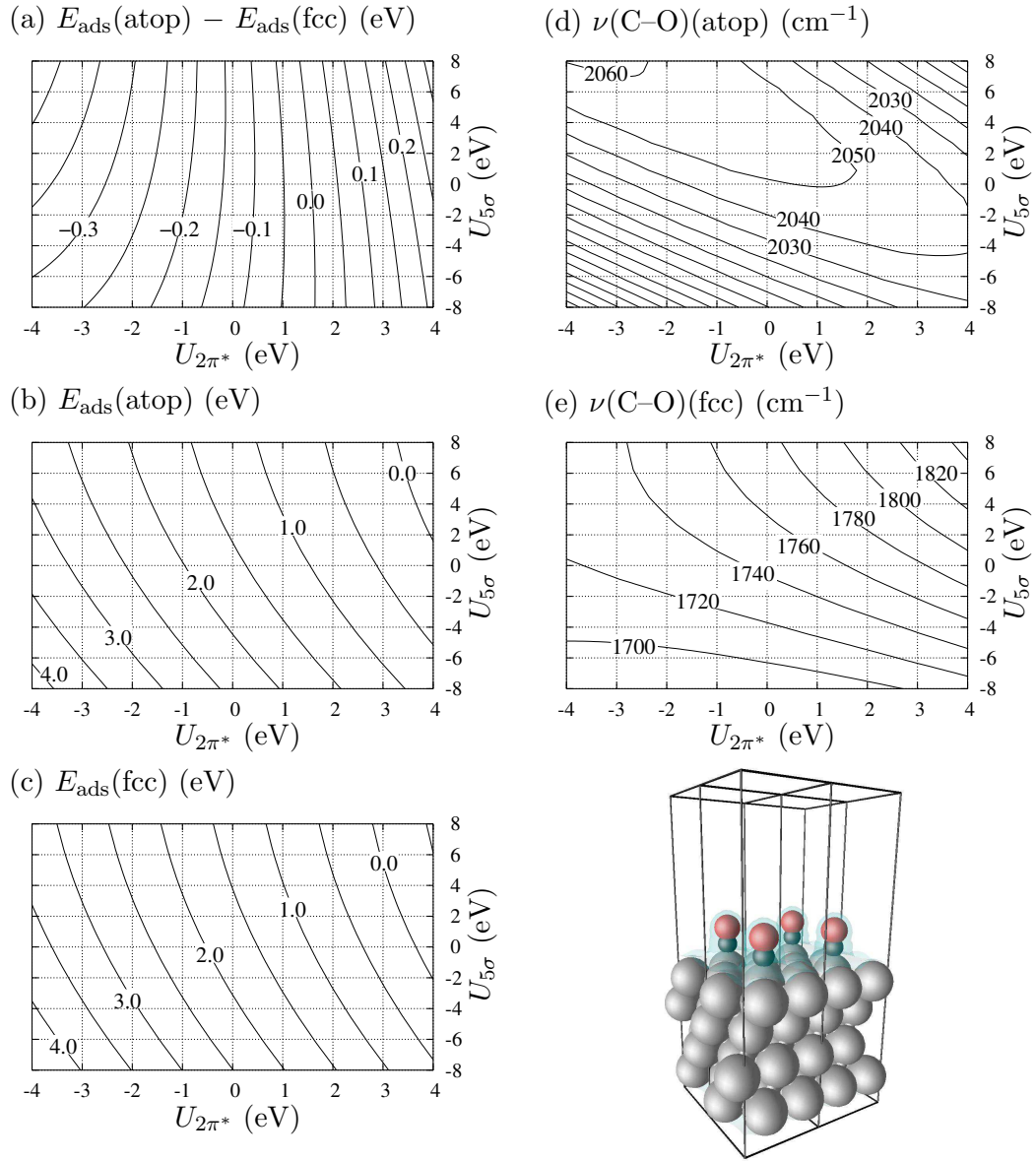}
\caption[Adsorption energy and stretching frequency
as a function of the hybridization parameters for CO
on platinum]{
Adsorption energy and intramolecular stretching frequency as a 
function of $U_{2\pi^*}$ and $U_{5\sigma}$ controlling the hybridization
of the LUMO and HOMO orbitals for atop and fcc adsorption
of CO on Pt(111).
\label{MolecularUResults}}
\end{center}
\end{figure}

The results of the calculations are presented in Figure \ref{MolecularUResults},
and in Appendix \ref{MolecularUFDOSAppendix}, along with methodological details.
As expected, adsorption energies decrease with increasing penalization on the
hybridizations of the
HOMO and LUMOs. Consequently, both donation and backdonation favor CO
adsorption, in agreement
with Ref. \cite{KresseGil2003}. Moreover, we observe that electron backdonation tends to
decrease the relative
adsorption energy $E_{\rm ads}({\rm atop})-E_{\rm ads}({\rm fcc})$, confirming 
that $2\pi^*$ backdonation favors
CO adsorption at
high-coordination sites, as demonstrated by Anderson and Awad
\cite{AndersonAwad1985}. Additionally,
the effect of $U_{5\sigma}$ on
the relative binding energy is much weaker than that of $U_{2\pi^*}$. This result
supports the hypothesis that
the failure of density-functional calculations in predicting the most stable
adsorption site is principally
related to an overestimation of $2\pi^*$ backdonation \cite{GajdosHafner2005,
KohlerKresse2004,KresseGil2003}.

Considering now the structural and vibrational properties, we observe more
complex $U_{2\pi^*}$-  and $U_{5\sigma}$-dependencies. 
The effect of electron donation must clearly be taken into account
when analyzing the
sensitivity of the calculated stretching frequencies. Note that the invariance
of the stretching frequency
with respect to $U_{2\pi^*}$ for CO adsorbed at the atop site, as already observed in
Ref. \cite{GajdosHafner2005} for copper
surfaces, can be explained by the fact that the $U_{2\pi^*}$-axis is tangent to the
contour line $\nu(\textrm{C--O})$ = 2050
cm$^{-1}$. Despite this fact, the dependence of $\nu(\textrm{C--O})(\rm atop)$ 
with respect to $U_{5\sigma}$ is
appreciable, supporting
the idea that $2\pi^*$ backdonation alone does not control the site-dependence of
the C--O stretching
frequency.

\begin{figure}
\centering
\raisebox{2.5cm}{
\begin{minipage}[t]{0.38\linewidth}
\caption[Orbital-hybridization sensitivity analysis fo
the adsorption energy, bond length, and stretching frequency of
CO on platinum]{
Ranges of variation (indicated by black error bars) of
the adsorption energy, bond length, and stretching frequency of
CO on Pt(111) for a very broad range of hybridizations
(0 eV $<$ $U_{2\pi^*}$ $<$ 5 eV and $-10$ eV $<$ $U_{5\sigma}$ $<$ 10 eV).
\label{Sensitivity}
}
\end{minipage}}\hfill
\begin{minipage}[c]{.58\linewidth}
\includegraphics[width=8.75cm]{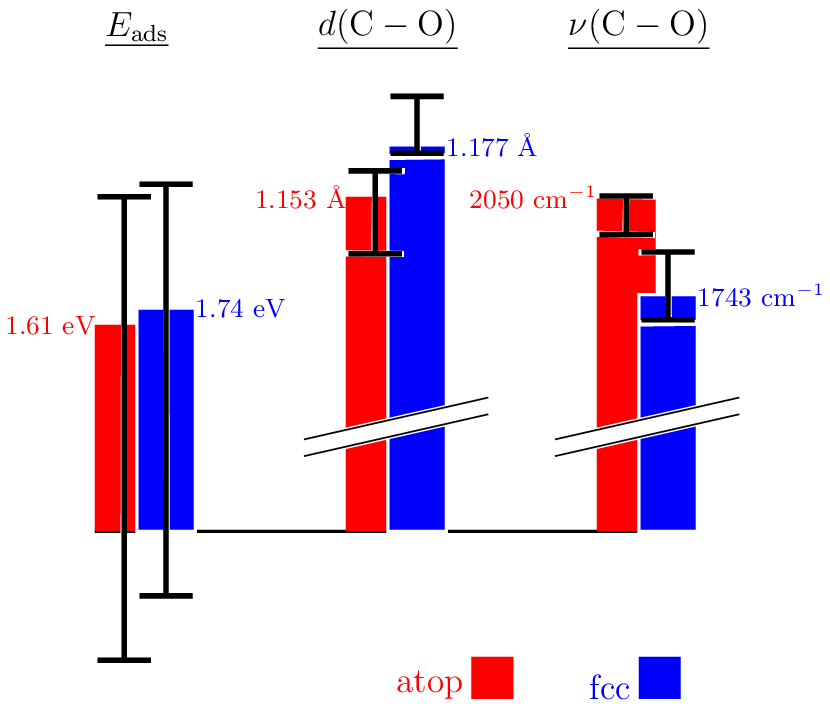}
\end{minipage}
\end{figure}

To conclude this section, we report the results of our sensitivity analysis
(Figure \ref{Sensitivity}). Large ranges for
$U_{2\pi^*}$and $U_{5\sigma}$ are selected: 0 eV $<$ $U_{2\pi^*}$ $<$ 5 ev and 
$-10$ eV $<$ $U_{5\sigma}$ $<$ 10 eV.
Note that the parameter
$U_{2\pi^*}$ is kept positive since the GGA + molecular U calculations clearly
indicate that $2\pi^*$ backdonation
is overestimated. We emphasize that these energy ranges correspond to large
shifts in the single-electron
energies (up to $\Delta \epsilon_{2\pi^*}$ = 1 eV and 
$|\Delta \epsilon_{5\sigma}|$ = 4 eV) and to large variations of
the adsorption
energies (up to $|\Delta E_{\rm ads}|$ = 1.5 eV). Thus, the relative variations of the
adsorption energies are
comparable to their absolute values. Despite these sizable variations of the
adsorption energies, we
observe little variations of the bond lengths and stretching frequencies:
\begin{equation}
\left\{ \begin{array}{lcccl}
1.127 \textrm{ \AA} & < & d{\rm (C-O)(atop)} & < &  1.165 \textrm{ \AA} \\
1.173 \textrm{ \AA} & < & d{\rm (C-O)(fcc)} & < &  1.199 \textrm{ \AA} \\
1933 \textrm{ cm}^{-1} & < & \nu{\rm (C-O)(atop)} & < & 2055 \textrm{ cm}^{-1} \\
1666 \textrm{ cm}^{-1} & < & \nu{\rm (C-O)(fcc)} & < & 1878 \textrm{ cm}^{-1} 
\end{array}
\right.
\end{equation}

These small variations account for the remarkable accuracy of the bond lengths
and stretching
frequencies calculated within PBE-GGA. In particular, they justify the correct
ordering of the C--O
stretching frequencies despite important qualitative errors in predicting the
relative CO adsorption
energies. These results provide strong support to the conclusion of the
preceding section: the variations
of $d(\textrm{C--O})$ and $\nu(\textrm{C--O})$ are not directly due to 
electron donation and backdonation,
but rather to the
hybridization of the $1\pi$ orbitals.

\section{Conclusion}

This study has evidenced that the PBE-GGA predictions for the stretching
frequencies of CO
adsorbed on platinum are in excellent agreement with SFG experiments despite the
well-known failure
of local and generalized-gradient calculations in predicting the most stable
adsorption site. Similar
agreement is obtained for CO adsorbed on platinum-ruthenium bimetallic surfaces,
allowing the direct
recognition of CO adsorption sites.

Our orbital-resolved force analysis has demonstrated that the variations of bond
length and stretching
frequency as a function of the CO adsorption site are principally due to the $1\pi$
hybridization, rather than
the $2\pi^*$ and $5\sigma$ hybridizations. Using the GGA + molecular U approach, we have
performed a
sensitivity analysis to quantify the influence of the $2\pi^*$ and $5\sigma$ hybridizations
on the structural and
vibrational properties for CO on platinum. The effect of $2\pi^*$ backdonation has
been shown to be small
and comparable to that of $5\sigma$ donation, contradicting the widespread idea that
backdonation controls
the frequency shifts.

These results explain the remarkable accuracy of the PBE-GGA frequency
predictions despite errors
in the hybridizations of the $2\pi^*$ and $5\sigma$ orbitals. Furthermore, they suggest a
promising way to connect
density-functional calculations with experiments in some of the most problematic
cases of molecular
adsorption on transition metals.

\begin{acknowledgments}
The calculations in this work have been performed using the 
Quantum-Espresso package (GNU General Public License).
The authors acknowledge support from the MURI grant DAAD 19-03-1-0169. 
I. D. personally thanks the \'Ecole Nationale des Ponts et Chauss\'ees (France) 
and the Martin Family Society of Fellows for Sustainability for their help and support. 
Valuable discussions with Matteo Cococcioni, Cody Friesen, and Fabien Sorin are gratefully
acknowledged.
\end{acknowledgments}

\appendix

\section{Ultrasoft Overlaps}

\label{UltrasoftAppendix}

The Vanderbilt ultrasoft formalism \cite{Vanderbilt1990} 
consists of replacing the density operator $\hat n({\bf r})$ with:
\begin{equation}
\hat n^{US}({\bf r}) = \hat n ({\bf r}) + \sum_I \hat n_I({\bf r})
\end{equation}
where $\hat n_I({\bf r};{\bf r}_1,{\bf r}_2) = \sum_{n,m} Q_{I,n,m}({\bf r})
\beta_{I,n}^*({\bf r}_1) \beta_{I,m}({\bf r}_2)$
is the charge-augmentation contribution from the ionic
core $I$. Correspondingly, the overlap operator $\hat S$ becomes:
\begin{equation}
\hat S = \hat 1 + \sum_I \hat S_I
\end{equation}
where $\hat S_I({\bf r}_1,{\bf r}_2) = \sum_{n,m} \int Q_{I,n,m}({\bf r}) d{\bf r}
\beta_{I,n}^*({\bf r}_1) \beta_{I,m}({\bf r}_2)$
is the ionic contribution to the overlap operator. The
ultrasoft pseudopotential of the ionic core $I$ is the sum of a local part 
$v_I^{L}({\bf r})$ and a nonlocal part
$\hat v_I^{NL}({\bf r}_1,{\bf r}_2) = \sum_{n,m}
D^0_{I,n,m} \beta_{I,n}^*({\bf r}_1) \beta_{I,m}({\bf r}_2)$.

This ultrasoft formalism considerably improves the convergence of 
density-functional algorithms with
respect to the energy cutoffs applied to the plane-wave expansions 
of the wavefunctions and charge
density. However, to calculate the overlap $\langle \chi |
\hat S | \psi \rangle$ between the molecular orbital $\chi$ and the
wavefunction $\psi$,
it must be borne in mind that the overlap operators $\hat S_\chi$ and $\hat S_\psi$ 
corresponding to $\chi$ and
$\psi$ are distinct since the molecular orbital is calculated without the platinum slab. 
The procedure
employed here consists in including fictitious platinum cores in the calculation of the molecular orbital.
These fictitious cores are obtained by setting the local and nonlocal part of the platinum pseudopotential
to zero, while keeping the contribution to the overlap operator unchanged. 
The resulting operator $\hat S^{fict}_\chi$
being identical to $\hat S_\psi$, the overlap coefficient can be calculated as 
$\langle \chi^{fict} | \hat S^{fict}_\chi | \psi \rangle = \langle \chi^{fict} | \hat S_\psi | \psi \rangle$ where
$\chi^{fict}$ is the molecular orbital calculated in the presence of the fictitious platinum cores. The primary
advantage of this procedure is that it only requires changing the pseudopotentials.

\section{Density-distribution Analysis}

\label{ZupanAnalysisAppendix}

\begin{figure}
\centering
\raisebox{2.7cm}{
\begin{minipage}[t]{0.38\linewidth}
\caption[Derivative of the density and density-gradient distribution functions
for CO on platinum]{
Derivative of the density distribution function
$\Delta^{(\textrm{C--O})}g_{1,\chi}(r_s)$ and derivative of the density-gradient
distribution function $\Delta^{(\textrm{C--O})}g_{3,\chi}(s)$ along the
C--O stretching mode for each molecular orbital $\chi$.
\label{ZupanDensities}
}
\end{minipage}}\hfill
\begin{minipage}[c]{.57\linewidth}
\includegraphics[width=9.25cm]{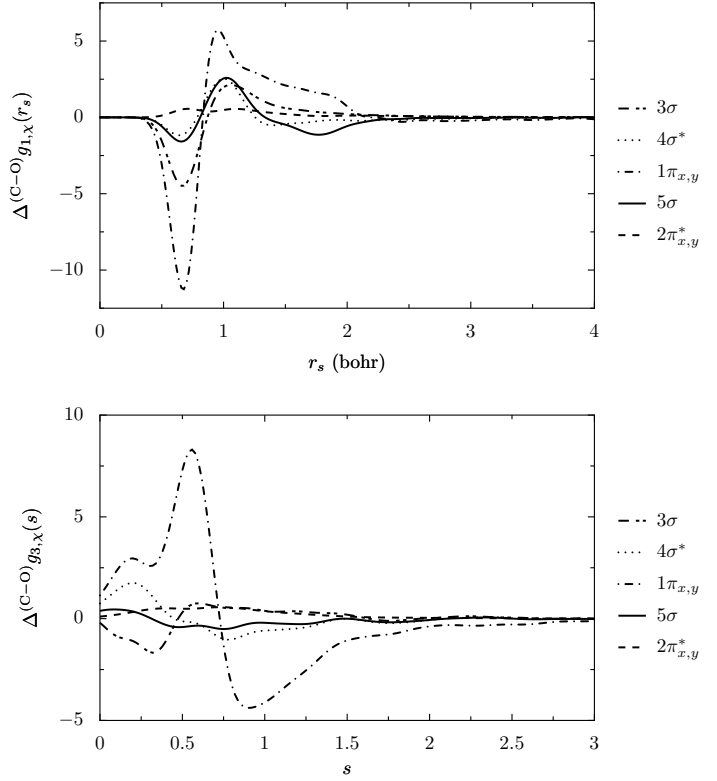}
\end{minipage}
\end{figure}

As an additional confirmation to the predominance of the $1\pi$ 
contribution in the C--O intramolecular
force, we have calculated the density distribution and density-gradient distribution
\cite{ZupanBurke1997} projected on each
CO molecular orbital $\chi$. We define the projected density distribution as:
\begin{equation}
g_{1,\chi}(r_s) = \sum_i f_i |\langle \chi | \psi_i \rangle |^2
\int |\psi_i({\bf r})|^2 \delta(r_s-r_s({\bf r})) d {\bf r},
\end{equation}
where $r_s$ is the Seitz radius. We propose a similar
 definition for the projected density-gradient
distribution:
\begin{equation}
g_{3,\chi}(s) = \sum_i f_i |\langle \chi | \psi_i \rangle |^2
\int |\psi_i({\bf r})|^2 \delta(s-s({\bf r})) d {\bf r},
\end{equation}
where $s = |\nabla n|/2k_Fn$ is the reduced density gradient. 
The derivatives of the distribution functions along
the C--O stretching mode $\Delta^{(\textrm{C--O})} g_{n,\chi}=
d(\textrm{C--O}) \partial g_{n,\chi} / \partial d(\textrm{C--O})$
($n$=1,3) are plotted in Figure \ref{ZupanDensities}.

Considering $\Delta^{(\textrm{C--O})}g_{1,\chi}(r_s)$, we observe that increasing 
$d(\textrm{C--O})$ tends to decrease the electronic charge
in spatial regions of high electronic density. 
This trend is particularly marked for the $1\pi$ molecular
orbitals, confirming their predominance in the intramolecular force. 
The predominant $1\pi$ contribution
can also be seen in the $\Delta^{(\textrm{C--O})}g_{3,\chi}(s)$ graph. 
It is important to note that the $1\pi$ orbitals result in a
significant increase in charge-density homogeneity. 
According to the bond-expansion criterion derived
by Zupan, Burke, Ernzerhof, and Perdew (Eq. 9 in Ref \cite{ZupanBurke1997}),
this observation confirms that the $1\pi$
orbitals are strongly bonding.

\section{GGA + molecular U Force Density of States}

\label{MolecularUFDOSAppendix}

\begin{figure}
\begin{center}
\includegraphics[height=20cm]{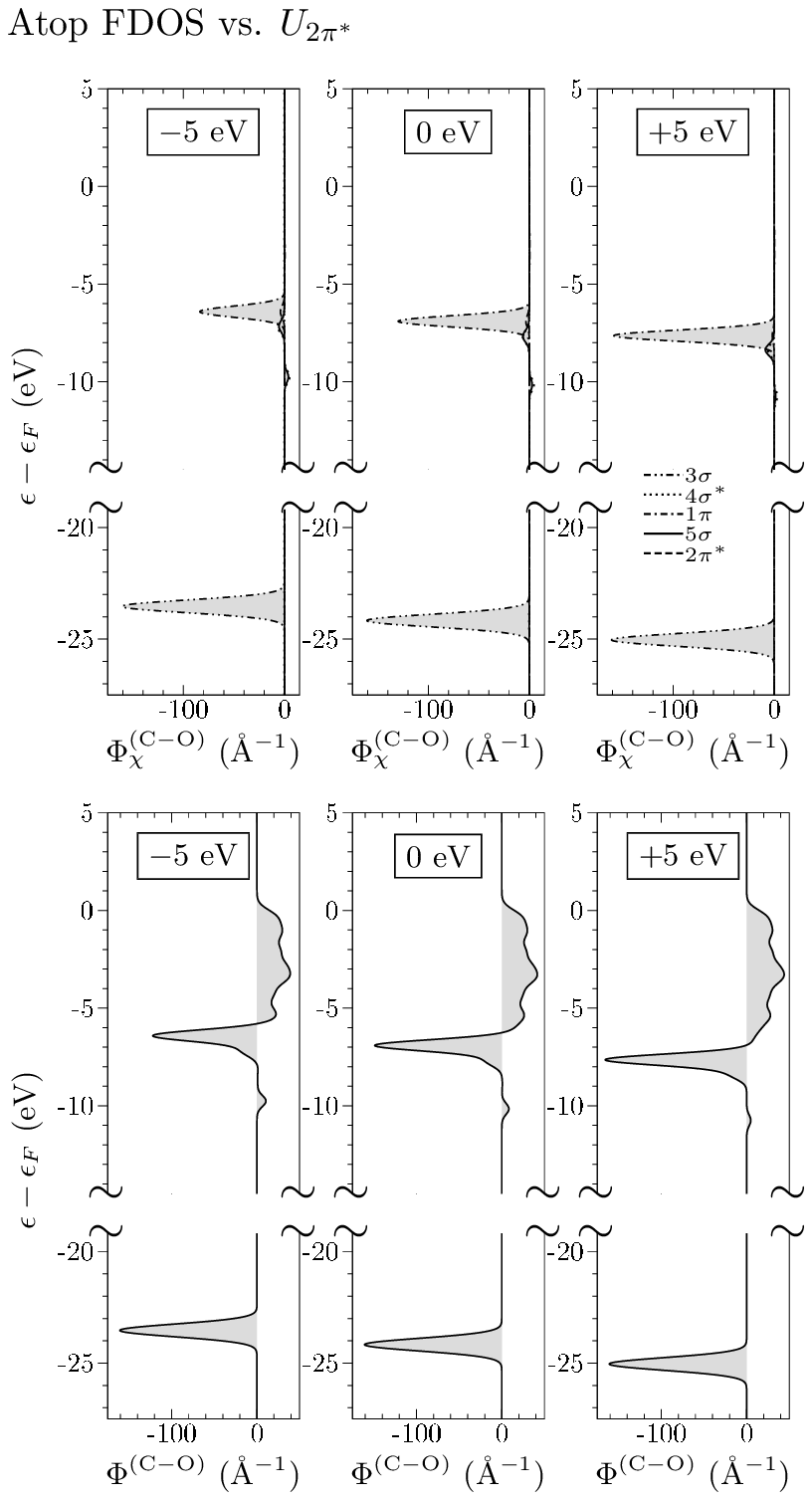}
\caption[Force density of states as a function of the
LUMO hybridization parameter for atop adsorption of CO on platinum]{
Force density of states projected on the CO molecular orbitals, and total force density of
states as functions of $U_{2\pi^*}$ 
controlling the hybridization of the LUMO orbitals for atop adsorption of CO on Pt(111).
\label{FigureUPiForce}}
\end{center}
\end{figure}

\begin{figure}
\begin{center}
\includegraphics[height=20cm]{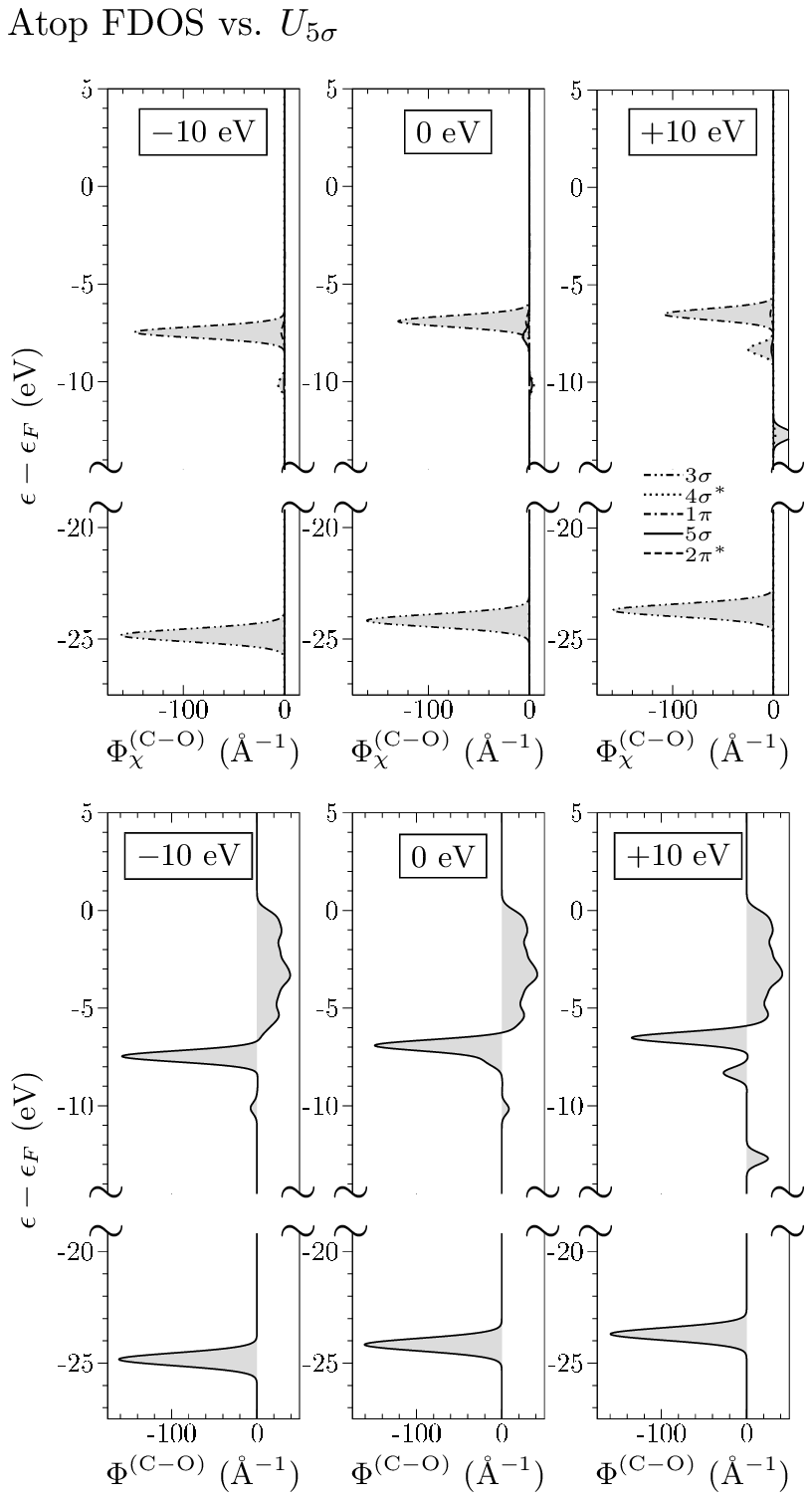}
\caption[Force density of states as a function of the
HOMO hybridization parameter for atop adsorption of CO on platinum]{
Force density of states projected on the CO molecular orbitals, and total force density of
states as functions of $U_{5\sigma}$ controlling the hybridization of the LUMO orbitals for atop adsorption of CO on Pt(111).
\label{FigureUSigmaForce}}
\end{center}
\end{figure}

The calculation of the FDOS within GGA + molecular U can be performed along the same general
lines as the method presented in Sec. \ref{ElectronicOriginSection}. 
the only modification being in the expression of the
wavefunction contribution to the force ${\bf F}_{I,i\sigma}$. 
In the ultrasoft formalism and in the absence of molecular U
contribution, the wavefunction contribution to the force can be written as
\cite{DalCorso2001,LaasonenPasquarello1993}:
\begin{equation}
\left( {\bf F}_{I,i\sigma} \right)_{GGA} = - f_{i\sigma} \langle \psi_{i\sigma} | 
\frac{\partial \hat v}{\partial {\bf R}_I} - \epsilon_{i\sigma} 
\frac{\partial \hat S}{\partial {\bf R}_I} | \psi_{i\sigma} \rangle.
\end{equation}
Adding the molecular U term, the force contribution becomes:
\begin{eqnarray}
\left( {\bf F}_{I,i\sigma} \right)_{GGA + U} & = & \left( {\bf F}_{I,i\sigma} \right)_{\rm GGA}
- \frac{U_{2 \pi^*}}{2} \sum_\sigma \textrm{Tr}\left\{ 
({\bf I} - 2 {\bf f}_{2 \pi^*, \sigma})\left.\frac{\partial {\bf f}_{2 \pi^*, i\sigma})}{\partial {\bf R}_I}\right|_{\psi_{i\sigma}}\right\} 
\nonumber \\
&  & - \frac{U_{5 \sigma}}{2} \sum_\sigma ( 1 -2 f_{5 \sigma, \sigma})
\left.\frac{\partial f_{5 \sigma, i\sigma}}{\partial {\bf R}_I}\right|_{\psi_{i\sigma}}
\end{eqnarray}
where ${\bf f}_{2 \pi^*, i\sigma}=[f_{i\sigma}
\langle \psi_{i\sigma}|\hat S|2\pi^*_\alpha \rangle \langle 2\pi^*_\beta|\hat S|
\psi_{i\sigma} \rangle]_{\alpha \beta}$ denotes the contribution from
 the wavefunction $\psi_{i\sigma}$
to the $2\pi^*$ occupation matrix ${\bf f}_{2 \pi^*, \sigma}$
and $f_{5 \sigma, i\sigma}=f_{i\sigma}
\langle \psi_{i\sigma}|\hat S| 5\sigma \rangle 
\langle 5\sigma|\hat S| \psi_{i\sigma} \rangle$
is the contribution from $\psi_{i\sigma}$ to
$f_{5 \sigma, \sigma}$.
Note that the derivatives must be calculated keeping $\psi_{i\sigma}$ fixed.

The main computational difficulty in determining the derivatives 
of the occupation coefficients is the
evaluation of the response of the molecular orbitals 
$\chi$ to the atomic displacements $\partial \chi/\partial {\bf R}_I$. These
responses can be obtained by performing a separate calculation for an isolated CO molecule using the
same linear-response approach as that employed in the phonon calculation 
\cite{BaronideGironcoli2001}.
The projected and total FDOS spectra as a function of the hybridization parameters 
for CO adsorbed at
the atop site on Pt(111) are reported in Figures \ref{FigureUPiForce} and 
\ref{FigureUSigmaForce}. 
The observed trends confirm that large shifts
in the LUMO and HOMO single-electron energies do not affect the predominance of the 
$1\pi$ contribution. For positive values of $U_{2 \pi^*}$ and 
$U_{5\sigma}$, the magnitude of the peak in the $1\pi$ FDOS increases
due to the stronger molecular character of the $1\pi$ orbitals. 
Note that this increase in the $1\pi$ bonding
contribution is accompanied by an increase in the antibonding contribution from the metal bands.
Similar compensation effects in the electronic forces can be observed for negative values of the
hybridization parameters. In particular, for $U_{2 \pi^*}$ equal to 
$-5$ eV corresponding to an unphysical
overestimation of the $2 \pi^*$ hybridization—the decrease 
in the bonding contribution from the $1\pi$ orbitals
is partially offset by the decrease in the antibonding contribution from the d bands. 
This observation
provides a more specific understanding of the invariance of the stretching frequency as a function of the HOMO and LUMO hybridizations.

\bibliography{article}

\end{document}